\documentclass[10pt,conference]{IEEEtran}
\newcommand{\tool}{\textsc{HarnessFix}\xspace}

\usepackage[T1]{fontenc}

\usepackage[linesnumbered,ruled,vlined]{algorithm2e}
\usepackage{algpseudocode}

\usepackage{flushend}
\usepackage{balance}
\usepackage{float}
\usepackage{wrapfig}
\usepackage{array}
\usepackage{tabularx}

\usepackage{threeparttable}
\usepackage{booktabs}
\usepackage{multirow}
\usepackage{amsmath}
\usepackage{amssymb}

\usepackage{graphicx}
\usepackage{epstopdf}
\usepackage{subcaption}
\usepackage{subfloat}
\usepackage{tikz}
\usepackage{pgfplots}
\usetikzlibrary{pgfplots.statistics,calc}

\usepackage{listings}
\usepackage{pythonhighlight}
\usepackage{minted}
\usepackage{verbatim}

\usepackage{hyperref}
\usepackage{cleveref}
\usepackage{url}

\usepackage[normalem]{ulem}
\usepackage{enumitem}
\usepackage{ragged2e}
\usepackage{xspace}
\usepackage{centernot}
\usepackage{CJKutf8}

\usepackage{marvosym}

\usepackage{bbding}
\usepackage{pifont}
\usepackage{wasysym}

\usepackage[table]{xcolor}
\usepackage{soul}

\usepackage{fancybox}
\usepackage[framemethod=TikZ]{mdframed}
\usepackage{tcolorbox}
\usepackage{caption}
\usepackage{xparse}

\newcommand\hlg{\bgroup\markoverwith
  {\textcolor{green!10}{\rule[-.5ex]{2pt}{2.5ex}}}\ULon}
\newcommand\hlb{\bgroup\markoverwith
  {\textcolor{blue!10}{\rule[-.5ex]{2pt}{2.5ex}}}\ULon}
\newcommand\hlr{\bgroup\markoverwith
  {\textcolor{red!10}{\rule[-.5ex]{2pt}{2.5ex}}}\ULon}

\newminted{c}{
    linenos,
    frame=lines,
    tabsize=4,
    autogobble,
}

\newminted{java}{
    linenos,
    frame=lines,
    tabsize=4,
    autogobble,
}

\newif\ifFirstMintedPart


\newcommand{\mybox}[1]{%
  \setbox0=\hbox{#1}%
  \setlength{\@tempdima}{\dimexpr\wd0+13pt}%
  \begin{tcolorbox}[boxrule=0.5pt, colback=white, arc=4pt,
      left=6pt,right=6pt,top=6pt,bottom=6pt,boxsep=0pt]
    #1
  \end{tcolorbox}
}

\def\BibTeX{{\rm B\kern-.05em{\sc i\kern-.025em b}\kern-.08em
    T\kern-.1667em\lower.7ex\hbox{E}\kern-.125emX}}

\begin{document}

\title{From Failed Trajectories to Reliable LLM Agents: Diagnosing and Repairing Harness Flaws}

\author{
Mengzhuo Chen$^{1,2,3}$,
Junjie Wang$^{1,2}$,
Zhe Liu$^{1,2}$,
Yawen Wang$^{1,2}$,
Haiming Zheng$^{4}$,
Qing Wang$^{1,2}$ \\
\\
$^{1}$State Key Laboratory of Complex System Modeling and Simulation Technology, Beijing, China \\
$^{2}$Institute of Software, Chinese Academy of Sciences, Beijing, China \\
$^{3}$University of Chinese Academy of Sciences, Beijing, China \\
$^{4}$School of Computer Science and Technology, Tianjin University, Tianjin, China \\
}

\maketitle

\begin{abstract}
LLM agents increasingly rely on agent harness: the runtime infrastructure around the base model that defines execution environments, tool interfaces, context, lifecycle orchestration, observability, verification, and governance.
Existing self-improving agents and automatic harness evolution methods mainly improve agents through runtime supervision, prompt optimization, workflow search, or harness modification based on final outcomes. However, they often fail to diagnose where the responsible evidence lies in failed trajectories and which harness implementation mechanism 
causes the unreliable behavior, resulting in broad, indirect, or poorly scoped changes.
This paper proposes {\tool}, a trace-grounded and
diagnosis-driven framework for repairing agent harnesses. {\tool} compiles raw execution traces and harness artifacts into a Harness-aware Trace Intermediate Representation (HTIR), which normalizes fragmented trajectory evidence, captures step-level data-flow and control-flow relations, and aligns runtime steps with the harness artifacts that shape their behavior.
It then attributes failures to responsible steps and harness artifacts, and consolidates recurring diagnoses into repair-oriented flaw records.
Finally, {\tool} maps these records to scoped repair operators, generates  patches under flaw-specific repair specifications, and accepts them through regression-aware validation.
We evaluate {\tool} on four popular benchmarks, and results show that it 
improves the performance over the initial harnesses by 6.3\% to 18.4\%, significantly outperforming human-designed and self-evolution baselines. 
{\tool} highlights the value of treating failed trajectories not only as feedback signals, but also as structured evidence for diagnosing and repairing the harness mechanisms behind agent failures.
\end{abstract}


\begin{table*}[!t]
\centering
\caption{Harness layers and their responsibilities~\cite{li2026agentharness}.}
\label{tab:htir}
\scriptsize
\rowcolors{2}{gray!6}{white}
\begin{tabular}{p{0.21\textwidth}p{0.74\textwidth}}
\toprule
\rowcolor{white}\textbf{Harness layer} & \textbf{Responsibilities} \\
\midrule
Execution Environment and Sandbox & Provides safe, isolated, reproducible environments in which agent actions run with bounded autonomy. \\
Tool Interface & Governs how agents discover, describe, select, and invoke tools, including schemas, documentation, and error feedback. \\
Context and Memory & Determines what the model sees at each step: context window, session state, summaries, retrieved evidence, and persistent memory. \\
Lifecycle and Orchestration & Controls agent execution flow: think-act-observe loops, retries, task-state management, multi-agent coordination, and termination. \\
Observability & Records traces, logs, tool calls, errors, and cost information well enough to diagnose failures and monitor behavior. \\
Verification and Evaluation & Connects tasks to feedback through readiness checks, intermediate validation, final output evaluation, and regression testing. \\
Governance and Security & Defines permissions, identities, policies, approvals, and audit trails that constrain agent authority. \\
\bottomrule
\end{tabular}
\rowcolors{2}{white}{white}
\end{table*}

\section{Introduction}
\label{sec:Introduction}
LLM agents are increasingly used for multi-step, tool-mediated tasks in repository-level software engineering, terminal-based workflows, open-ended research, and application automation~\cite{swebench2023,swebenchverified2024,terminalbench2026,gaia2023,appworld2024}. Unlike standalone LLM usage, these agents repeatedly interact with external environments through model invocations, tool calls, observations, intermediate artifacts, state changes, and final submissions. As a result, the reliability of such systems depends not only on the base model, but also on the agent harness: the runtime infrastructure that wraps the model and governs how it acts \cite{zhang2026agentdevelreframingselfevolvingllm,hao2026recreate,fang2026trajectory}.
Recent harness engineering work characterizes this infrastructure using the seven layer taxonomy, including Execution, Tooling, Context, Lifecycle, Observability, Verification, and Governance~\cite{li2026agentharness}.
For example, the tool interface determines how actions are exposed and invoked; context and memory determine what the model sees; and lifecycle and orchestration govern execution flow and task state.

When failures occur, improving the harness is often more practical than modifying the base model, which is typically expensive to retrain, fine-tune, or redeploy.
However, harness improvement is far from straightforward.
Unlike traditional software systems, where business logic is explicitly encoded in control flow and data structures \cite{zhang2025whichagent,bouzenia2025understandingsoftwareengineeringagents,islam2026agents}, the behavior of agents is largely determined at runtime through model inference, shaped by prompts, retrieved context, tool interfaces, and orchestration policies~\cite{li2026agentharness,openaiguideagents2025,anthropiccontext2025}.
Consequently, failures cannot usually be mapped directly to specific locations in the harness implementation, such as source code or prompts.
Instead, they emerge in execution trajectories that interleave natural language reasoning, tool interactions, environmental feedback, and intermediate state transitions.
The language-heavy, reasoning-driven, and non-deterministic nature of these trajectories makes traditional debugging and repair techniques substantially less effective \cite{zhang2013automated,chen2015deterministic,Monperrus18AutomaticSoftwareRepair,XiaWZ23AutomatedProgramRepair,JiangLLT23ImpactCodeLanguage}.

Existing efforts on self-improving agents and automatic harness evolution only partially address this problem.
One line of work focuses on runtime or supervisory optimization, using lightweight supervision and adaptive correction to bypass issues at execution time, suppress errors temporarily, or add external patches around the agent \cite{xiao2025improvingefficiencyllmagent,lin2025stopwastingtokensefficient,nanda2026wink,liu2026process}.
While such strategies can improve observed performance, they often do so without fixing the underlying harness flaw that caused the failure. Another line of work is primarily outcome-driven rather than diagnosis-driven: it optimizes prompts, workflows, or harness logic based on final scores, success rates, or efficiency metrics \cite{zhang2026agentdevelreframingselfevolvingllm,wang2025maestrojointgraph,costa2025instructionlevelweightshapingframework,scope2025,trace2skill2026}, without first identifying where the responsible evidence lies in the trajectory and which harness layer failed.
Without such targeted causal diagnosis, the resulting improvements are often broad, indirect, or poorly scoped.

In this work, we aim to start from failed agent trajectories, diagnose the responsible runtime steps in detail, and use this evidence to localize harness flaws and repair them correspondingly.
This goal raises several key challenges.
First, failure evidence is fragmented across language-heavy and reasoning-driven traces, spanning multi-step interactions, tool invocations, natural-language reasoning and environment responses, which makes precise diagnosis difficult.
Second, even if a failure can be localized within an execution trajectory, it remains difficult to map it back to the concrete harness implementation responsible for it, such as source code, prompt templates, tool specifications, configuration files, adapters, instrumentation hooks, and verification scripts.
This mapping is difficult because execution traces capture runtime behavior, whereas harness implementations are defined in static artifacts like code and prompts; in agent systems, these two views are often not explicitly aligned as in traditional software.
Third, harness modifications may reduce one recurring failure while introducing regressions elsewhere.
Although this challenge also appears in traditional program repair, it is more difficult in agent systems because the dependencies between runtime behavior and harness mechanisms are less explicit.

We propose {\tool}, a trace-guided framework for diagnosing agent failures and repairing agent harnesses.
It first abstracts raw execution traces and harness artifacts into a Harness-aware Trace Intermediate Representation (HTIR), which models step-level runtime behavior, reconstructs data-flow and control-flow links, and aligns the trajectory with the corresponding harness implementation artifacts.
HTIR provides the structured diagnostic basis for failure attribution by organizing fragmented trajectory evidence into attribution-ready step-level relations, enabling {\tool} to identify failure responsible runtime steps, explain root causes, and consolidate recurring diagnoses into flaw records.
For each flaw record, {\tool} then uses the runtime-to-implementation mapping captured in HTIR to localize repair targets, select scoped repair operators, and instantiate flaw-specific repair specifications for constrained patch generation and validation.

The repair operators used by {\tool} are grounded in our empirical study of real-world LLM agent development.
We analyze 30 popular open-source agent repositories and approximately 57,780 development records.
The study shows that harness-related changes are common across all repositories and that harness flaws span all seven ETCLOVG responsibilities, rather than being confined to prompts or a single component.
From recurring flaw patterns and developer repair practices, we summarize the scoped repair operators used by {\tool}.
These findings motivate {\tool}'s design choice to treat harness repair as a multi-layer runtime-infrastructure problem and to constrain generated patches with repair operators.

We evaluate {\tool} on four popular benchmarks: GAIA, SWE-Bench Verified, AppWorld, and Terminal-Bench 2.0 Verified.
Across these benchmarks, {\tool} improves performance over the initial harnesses by 6.3\%--18.4\%, and significantly outperforms human-designed and self-evolution baselines.
Ablation results further support the main design choices, showing that trace-grounded diagnosis and scoped repair are important to the final gains.

This paper makes the following contributions:
\begin{itemize}
    \item We present {\tool}\footnote{We release the source code on the anonymous website \url{https://github.com/HarnessFix/HarnessFix}. \label{github}}, a trace-grounded and diagnosis-driven framework for repairing agent harnesses. Unlike outcome-driven prompt evolution or free-form self-editing, {\tool} explicitly connects failure evidence to harness-aware diagnosis and scoped harness changes.
    \item We conduct the first empirical study of harness flaws in real-world LLM agents, to reveal flaw distributions across harness layers and summarize recurring repair operators that guide scoped repair in {\tool}.
    \item We introduce HTIR, a harness-aware trace representation that normalizes heterogeneous agent trajectories into step-level evidence structures, providing a basis for step-level attribution, harness diagnosis, and scoped repair planning.
    \item We evaluate {\tool} across four popular benchmarks and agents, showing that trace-grounded harness repair improves performance over initial harnesses by 6.3\%--18.4\%, and significantly outperforming human-designed harnesses and self-evolution baselines.
\end{itemize}

\section{Background and Motivation}
\label{sec_background}

\subsection{Background}
\label{sec:background}
A common perspective in agent engineering is that an LLM-based agent consists of two parts: the base model itself and everything else that surrounds it, i.e., the runtime infrastructure that provides tools, context, orchestration logic, etc. 
This surrounding infrastructure is collectively referred to as the \texttt{agent harness} \cite{li2026agentharness}. 
When failures occur, they often stem not from the model’s reasoning alone, but from how the harness exposes information, enables actions, records evidence, and enforces policies.

The harness is realized through editable implementation artifacts such as prompt templates, tool specifications, orchestration code, configuration files, adapters, logging hooks, and verification scripts.
Throughout this paper, \textit{harness repair} refers to modifying these artifacts, rather than updating the base-model parameters, to correct diagnosed flaws.

To describe the mechanisms that make up an agent harness, recent work introduces the ETCLOVG taxonomy~\cite{li2026agentharness}, which organizes harness responsibilities into seven layers as shown in Table~\ref{tab:htir}.
In this paper, we use these layers as a vocabulary for connecting runtime evidence to the harness mechanisms that may need repair.

\begin{table*}[!t]
\centering
\caption{Repair operators and associated flaws organized by harness layer
}
\label{tab:operators}
\footnotesize
\rowcolors{2}{gray!6}{white}
\begin{tabular}{>{\raggedright\arraybackslash}m{0.21\linewidth}>{\raggedright\arraybackslash}m{0.32\linewidth}>{\raggedright\arraybackslash}m{0.39\linewidth}}
\toprule
\rowcolor{white}\textbf{Harness layer} & \textbf{Typical flaws} & \textbf{Candidate repair operators} \\
\midrule
Execution Environment and Sandbox & Actions run in environments that are non-reproducible, insufficiently isolated, state-leaking, or blocked by interactive prompts. & Sandbox-boundary tightening; environment snapshot/restore; file-system/network/API access isolation; task-specific sandboxing; environment and state-diff logging. \\
Tool Interface & Tool schemas are ambiguous, candidate tool sets are poorly ranked, parameters are weakly validated, or tool errors are not actionable. & Tool-schema narrowing; argument validation; tool-candidate ranking and retrieval; tool documentation and error-message repair; read/write tool separation; transactional tool wrapping. \\
Context and Memory & Critical evidence, constraints, state, or prior results are omitted, stale, polluted, overly long, or fragmented. & Failure-relevant evidence preservation; artifact/state summary exposure; task-constraint refresh; summary and retrieval policy repair; persistent-memory isolation; context-budget checking. \\
Lifecycle and Orchestration & The control flow enters retry loops, repeats ineffective actions, finalizes prematurely, loses task state, or delegates without verification. & Loop guarding; explicit task-state modeling; retry and timeout bounding; verification-gated finalization; workflow-state checkpointing; delegated-output validation. \\
Observability & Traces lack the fields needed to diagnose failures or connect outcomes to requests, tool calls, errors, and state changes. & Model-request instrumentation; context-snapshot logging; tool-call/result tracing; API-error and state-delta logging; retry/token/latency tracking; structured trace logging. \\
Verification and Evaluation & Readiness checks, validators, judges, or regression tests do not catch invalid artifacts, missing effects, or task-inconsistent states. & Pre-execution readiness checking; intermediate validation gating; expected/actual state comparison; effect-evidence completion guarding; finalization-check strengthening; regression testing. \\
Governance and Security & Permissions, policy checks, approval flows, identity boundaries, and audit trails are too weak for high-impact actions. & Least-privilege credentialing; high-impact action approval gating; policy-check enforcement; audit-decision logging; out-of-scope action blocking; escalation-rule definition. \\
\bottomrule
\end{tabular}
\rowcolors{2}{white}{white}
\end{table*}

\subsection{Motivational Study}
\label{sec:motivation}

To understand how harness flaws arise and are addressed during the evolution of real-world LLM agent systems, we conduct a motivational study on popular open-source agents. 

\subsubsection{\textbf{Data Collection}}
\label{sec:motivation_collection}
We collect popular and actively open-source LLM agents from GitHub across diverse task categories, yielding 30 LLM agent systems and approximately 57,780 development records, including issues, pull requests, commits, and release notes. 
We classify development records as harness-related or unrelated using an LLM-based analysis agent with human validation through an iterative process, including small-scale annotation for examples, agent-based classification, manual inspection, agent strategy refinement, and a final sampled manual audit; due to space limitations, detailed collection and classification procedures are provided on our website.
The resulting dataset contains 26,174 harness-related records.

Overall, harness-related records account for approximately 45.3\% of all collected development records, and all 30 repositories contain such changes, indicating that developers frequently encounter harness-related issues in practice and highlighting the need for automated techniques to support harness maintenance and repair.

We further use a similar LLM-assisted and human-validated process to characterize the distribution of harness flaws and summarize common repair strategies, with detailed procedures provided on our website due to space limitations.

\begin{figure}[htb]
\centering
\includegraphics[width=8.7cm]{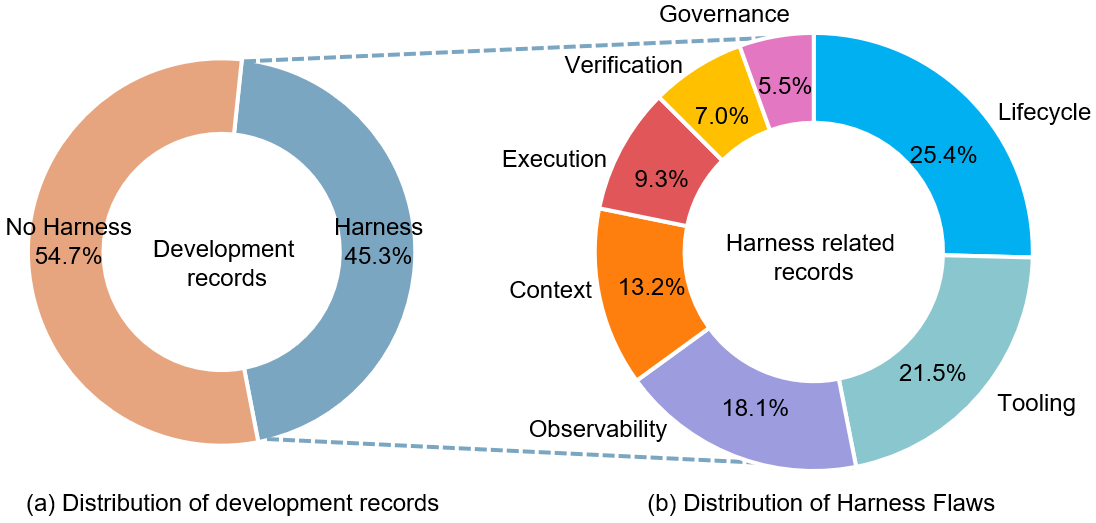}
\caption{Distribution across ETCLOVG layers.}
\label{fig:distribution}
\end{figure}

\subsubsection{\textbf{Distribution of Harness Flaws}}
\label{sec:motivation_flaws}

Figure \ref{fig:distribution} shows the distribution of harness flaws across the seven harness layers.
Flaws appear in all seven layers, and Lifecycle, Tooling, and Observability are the most frequent sources of flaws, with nearly all LLM agents in our dataset (29 out of 30) exhibiting flaws related to these three layers. 
We also observe that the dominant flaw layer varies with repository focus. Lifecycle flaws are more common in long-running or long-horizon agent systems~\cite{anthropicharnesses2025,horizon2026,miroflow2025,deepresearchagent2025,opendeepresearch2025,sweagent2024,harborterminus2026}, while
verification flaws are more frequently observed in benchmark-facing harnesses and repositories~\cite{sweagent2024,swebench2023,swebenchverified2024,gaia2023,appworld2024,terminalbench2026,ma2026maestromultiagentevaluationsuite,guo2026agenteval}.
These results show that harness flaws span the runtime infrastructure rather than being limited to prompts or a single component, highlighting the need for repair techniques that can address multiple harness layers, beyond the prompt optimization or runtime-context adjustment targeted by existing techniques \cite{agrawal2026gepa,scope2025,xiao2025improvingefficiencyllmagent,lin2025stopwastingtokensefficient}.

\subsubsection{\textbf{Repair Strategies}}
\label{sec:motivation_operators}
Table~\ref{tab:operators} lists the typical repair strategies we summarize for each harness layer. We organize these strategies into a compact set of repair operators, each corresponding to a common class of harness repair actions.
These operators are used in Section~\ref{subsec:scoped_repair} to guide scoped harness repair. By constraining repair generation, they reduce the risk of unstable edits or broad regressions caused by free-form modification techniques to harness artifacts~\cite{godelagent2024,sica2025,dgm2025,huxleygodel2026}.

\begin{figure*}[t]
\centering
\includegraphics[width=\textwidth]{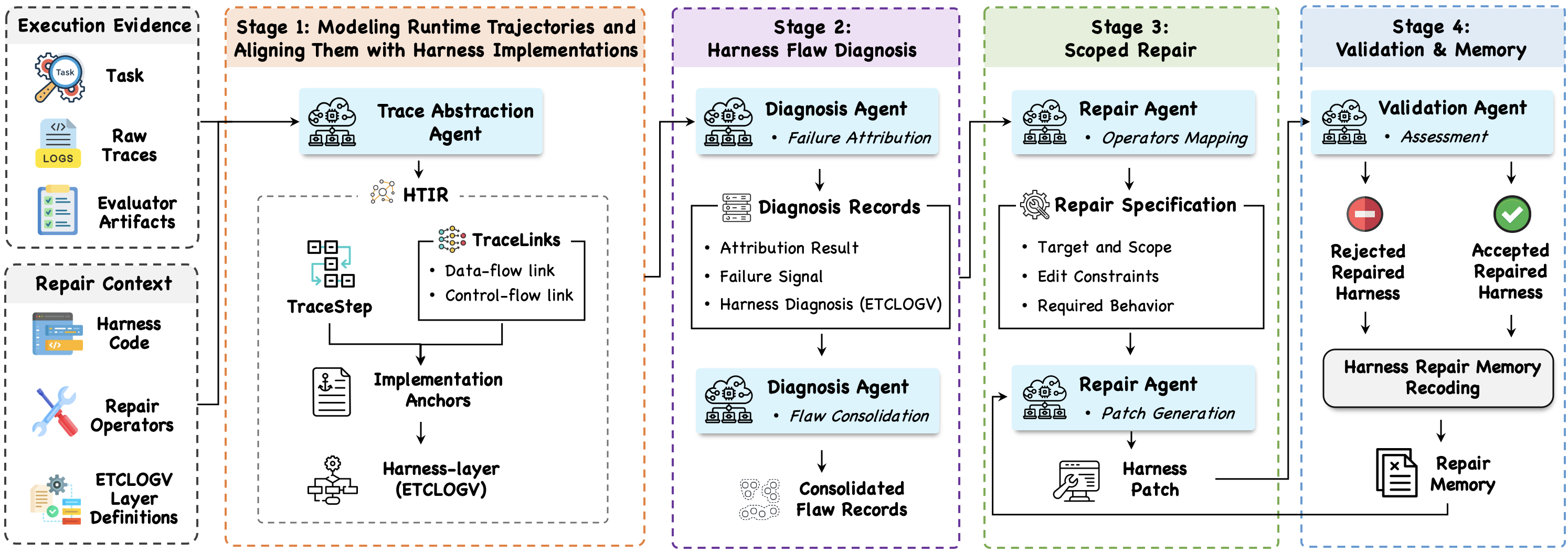}
\caption{Overview of {\tool}.}
\label{fig:overview}
\end{figure*}

\section{Approach}
\label{sec_approach}
The key idea of {\tool} is to drive harness repair from fine-grained, trace-grounded failure diagnoses.
It first localizes responsible runtime steps, maps their behavior to implicated harness flaws, and then translates those diagnoses into scoped repair specifications for guided repair.
{\tool} uses four LLM agents for trace-guided harness repair.
The trace abstraction agent constructs a Harness-aware Trace Intermediate Representation (HTIR) from raw traces and harness artifacts by modeling runtime behavior as TraceSteps, reconstructing data-flow and control-flow links, and aligning the resulting runtime evidence with editable harness artifacts.
The diagnosis agent consumes HTIR to attribute failures to responsible TraceSteps and implicated harness layers, and consolidates similar diagnoses into flaw records that summarize recurring root causes.
The repair agent maps flaw records to repair operators, instantiates flaw-specific repair specifications, and generates scoped candidate patches accordingly.
The validation agent checks whether the implementation stays within the intended repair scope and reduces the target flaw without introducing unacceptable regressions.

\begin{figure*}[t]
\centering
\includegraphics[width=\textwidth]{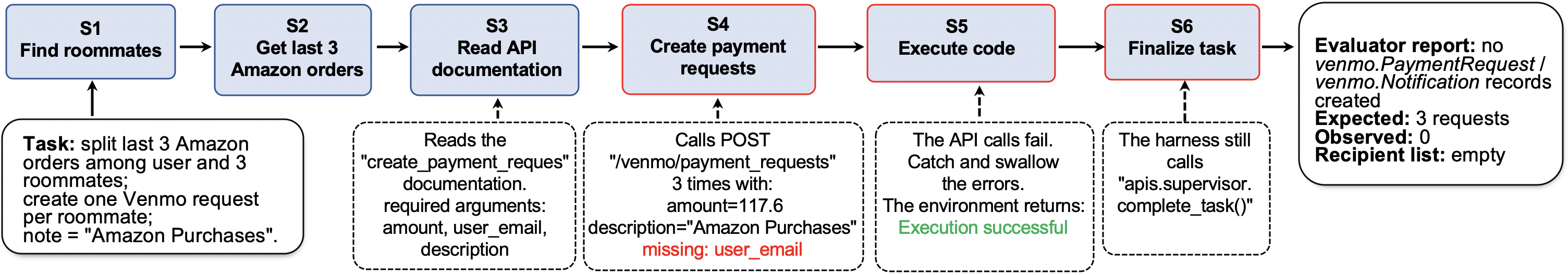}
\caption{Illustrative example for the AppWorld completion-guard harness flaw.
}
\label{fig:htir_example}
\end{figure*}

Before introducing {\tool}, we first present a running example from our experimental benchmark AppWorld, shown in \Cref{fig:htir_example}, which we use throughout the rest of this section to illustrate the approach.

\subsection{Modeling Runtime Trajectories and Aligning Them with Harness Implementations}

A raw trajectory records runtime behavior, but harness repair requires both identifying the runtime evidence responsible for a failure and locating the editable harness artifacts that realize or govern that evidence.
{\tool} therefore first constructs a Harness-aware Trace Intermediate Representation (HTIR), which serves two purposes.
First, it models the trajectory as TraceSteps and reconstructs cross-step TraceLinks, including data-flow and control-flow links, which provide the evidence needed for step-level failure attribution.
Second, it aligns TraceSteps and TraceLinks with harness implementations through implementation anchors, enabling responsible runtime evidence to be mapped to concrete artifacts.
This modeling and alignment stage provides the basis for identifying responsible runtime steps, diagnosing implicated ETCLOVG layers, and constraining later harness repair.

\subsubsection{\textbf{TraceStep: Unit of Runtime Evidence}}
Each node in HTIR represents a recoverable execution step from the trace, such as a model call, tool-mediated action, or final submission decision. We refer to each node as a \textit{TraceStep}.

Each TraceStep is assigned a unique identifier, which is a sequential integer indicating its position in the execution order. It also stores two basic fields, namely \textit{request message} and \textit{response message}. They are respectively the full message sent to or returned by the model, tool, or environment at this step.
HTIR also adds three derived annotations to each TraceStep. (a)\textit{ Role} is the step's function, such as information acquisition, tool invocation, artifact editing, validation, orchestration decision, or final submission. (b)\textit{ Execution status} includes success, failure, timeout, blocked, etc.
(c)\textit{ Artifact/state effect} records the type of externally observable effect produced by the step, together with corresponding effect details. Its effect type indicates whether the step has no observable external effect, a read-only interaction, an artifact change, an environment or application state change, a mixed effect, or an unknown effect.

The trace abstraction agent examines each TraceStep's request and response messages to infer these three annotations.
For artifact/state effect, the agent inspects the messages along with any available tool or environment metadata to identify externally observable consequences.
The effect details specify three kinds of information when available: a) the affected artifact or state entity, such as a file, database record, or session state; b) the observed state transition, such as file creation, database update, or session-state change;
and c) supporting evidence, such as tool return values, execution logs, or state-diff snapshots. The example below shows a TraceStep with its request, response, and derived annotations.

\begin{figure}[H]
\centering
\scriptsize
\setlength{\fboxsep}{1pt}
\begin{tikzpicture}[
    msg/.style={draw, rounded corners=1pt, align=left, inner sep=3pt, text width=0.43\columnwidth, font=\scriptsize},
    ann/.style={draw, rounded corners=1pt, align=left, inner sep=3pt, text width=0.39\columnwidth, font=\scriptsize}
]
\node[msg] (m) at (0.22\columnwidth,0) {\textbf{TraceStep S4}\\[-1pt]
{\tiny$\bullet$}~\texttt{id=S4}\\
{\tiny$\bullet$}~\texttt{request=POST /venmo/...}\\
\hspace{0.85em}\texttt{body=\{amount, description\}}\\
{\tiny$\bullet$}~\texttt{response=API call caught}};
\node[ann] (a) at (0.77\columnwidth,0) {\textbf{Derived annotations}\\[-1pt]
{\tiny$\bullet$}~\texttt{role=tool action}\\
{\tiny$\bullet$}~\texttt{status=success}\\
{\tiny$\bullet$}~\texttt{artifact/state effect:}\\[-1pt]
\hspace{0.85em}\texttt{no observable}\\[-1pt]
\hspace{0.85em}\texttt{external effect}\\
{\tiny$\bullet$}~\texttt{...}};
\draw[->, thick, gray!70] (m.east) -- node[above]{derive} (a.west);
\end{tikzpicture}
\end{figure}

\subsubsection{\textbf{Data-flow Alignment: Linking Evidence Propagation to Harness Logic}}

{\tool} constructs data-flow links to explain how information entered, disappeared from, or was transformed in a later model-facing request through earlier trajectory content and request construction logic.
This alignment is essential because many harness flaws arise when critical evidence, constraints, tool outputs, error messages, or state changes are lost, stale, polluted, incorrectly summarized, or not consumed by later steps.

To derive data-flow links, the trace abstraction agent jointly inspects the current request message, earlier TraceSteps, and the harness artifacts that assembles model-facing inputs.
It searches earlier TraceSteps in reverse temporal order and identifies both explicit reuse, such as copied or concatenated message spans, and semantic reuse, where the current request is grounded in earlier observations despite different surface wording.
It also inspects request construction logic, including memory assembly, tool descriptions, prompt templates, and other prompt building components, to determine how evidence was inserted, summarized, filtered, or omitted.

Each data-flow link records the source TraceStep id, target TraceStep id, source span in the earlier request or response, target span in the current request, and the reuse relation, such as copied, summarized, or semantically reused.
The span is a precise slice of the original trace record that verifiably grounds the link, enabling concrete evidence attribution.

The figure below shows an example data-flow link.
\begin{figure}[H]
\centering
\scriptsize
\setlength{\fboxsep}{1pt}
\begin{tikzpicture}[
    rec/.style={draw, rounded corners=1pt, align=left, inner sep=3pt, text width=0.92\columnwidth, font=\scriptsize}
]
\node[rec] (a) at (0.48\columnwidth,0) {\textbf{Data-flow link: S3 $\rightarrow$ S4}\\[-1pt]
{\tiny$\bullet$}~\texttt{source id: S3 (API documentation)}\\
{\tiny$\bullet$}~\texttt{target id: S4 (payment request)}\\
{\tiny$\bullet$}~\texttt{source span:}\\[-1pt]
\colorbox{yellow!25}{\texttt{... create\_payment\_request API docs ...}}\\[-1pt]
\colorbox{yellow!25}{\texttt{args: amount, description, user\_email, ...}}\\[-1pt]
\colorbox{yellow!25}{\texttt{user\_email: email of payment receiver}}\\[-1pt]
\colorbox{yellow!25}{\texttt{required fields include user\_email ...}}\\[-1pt]
{\tiny$\bullet$}~\texttt{target span:}\\[-1pt]
\colorbox{red!10}{\texttt{... create\_payment\_request(amount=54.0,}}\\[-1pt]
\colorbox{red!10}{\texttt{description=Amazon Purchases, ... )}}\\[-1pt]
\colorbox{red!10}{\texttt{body=\{amount, description, ...\}}}\\[-1pt]
{\tiny$\bullet$}~\texttt{reuse relation: semantically reused}};

\end{tikzpicture}
\end{figure}

The S3 API documentation for \texttt{create\_payment\_request}, including its required arguments and request schema, is semantically linked to the S4 payment request body.
This link makes the missing required \texttt{user\_email} field observable, thereby supporting the subsequent failure diagnosis.

\subsubsection{\textbf{Control-flow Alignment: Explaining Harness-driven Execution Decisions}}

{\tool} constructs control-flow links to explain why the harness transitions from one TraceStep to another.
Data-flow and control-flow links are complementary because agent failures may originate from either cross-step information propagation or step-transition logic.

To derive control-flow links, the trace abstraction agent examines the current TraceStep, the preceding and following TraceSteps in its temporal neighborhood, associated data-flow links, and the relevant harness implementations. It then matches the observed transition against harness execution logic, such as continue, retry, delegate, validate, finalize, or terminate, and records the condition or execution status that triggered the current TraceStep when such evidence is available.
Each control-flow link records the source TraceStep id, target TraceStep id, triggering logic, and the triggering condition or execution status when available.

The figure below shows an example control-flow link.
Although S5 returns \texttt{success}, its artifact/state effect is \texttt{no observable external effect}, showing that the expected state change has not occurred.
The completion guard incorrectly treats this execution status (i.e., success) as evidence of task progress and allows S6, a finalization TraceStep, to invoke \texttt{complete\_task()}.
This link therefore captures the controller decision that leads to incorrect finalization.

\begin{figure}[H]
\refstepcounter{figure}\label{fig:controlflow_message_link}
\centering
\scriptsize
\setlength{\fboxsep}{1pt}
\begin{tikzpicture}[
    tracebox/.style={draw=black, fill=white, rounded corners=1pt, align=left, inner sep=3pt, text width=0.39\columnwidth, font=\scriptsize},
    linkbox/.style={draw=black, rounded corners=1pt, align=left, inner sep=3pt, text width=0.50\columnwidth, font=\scriptsize, text=black},
    implbox/.style={draw, dashed, rounded corners=1pt, align=left, inner sep=3pt, text width=0.50\columnwidth, font=\scriptsize, text=gray!80},
    arr/.style={->, thick, black},
    dasharr/.style={->, dashed, gray!70}
]
\node[tracebox] (steps) at (0.20\columnwidth,-0.32) {\textbf{Source TraceStep S5}\\[-1pt]
{\tiny$\bullet$}~\texttt{role=tool action}\\
{\tiny$\bullet$}~\texttt{status=success}\\
{\tiny$\bullet$}~\texttt{artifact/state effect:}\\[-1pt]
\hspace{0.85em}\texttt{no observable}\\[-1pt]
\hspace{0.85em}\texttt{external effect}\\
{\tiny$\bullet$}~\texttt{...}\\[2pt]
\textbf{Target TraceStep S6}\\[-1pt]
{\tiny$\bullet$}~\texttt{role=finalization}\\
{\tiny$\bullet$}~\texttt{request=... complete\_task() ...}\\
{\tiny$\bullet$}~\texttt{status=success}\\
{\tiny$\bullet$}~\texttt{...}};
\node[linkbox] (cf) at (0.74\columnwidth,0.78) {\textbf{Control-flow link: S5 $\rightarrow$ S6}\\[-1pt]
{\tiny$\bullet$}~\texttt{source id: S5}\\
{\tiny$\bullet$}~\texttt{target id: S6}\\
{\tiny$\bullet$}~\texttt{triggering logic:}\\[-1pt]
\hspace{0.85em}\texttt{completion guard}\\
{\tiny$\bullet$}~\texttt{triggering condition:}\\[-1pt]
\hspace{0.85em}\texttt{status=success}};
\node[implbox] (impl) at (0.74\columnwidth,-1.78) {\textbf{Harness artifact reference}\\[-1pt]
\textit{Artifact location}\\[-1pt]
{\tiny$\bullet$}~\texttt{file: appworld\_agent/core.py}\\
{\tiny$\bullet$}~\texttt{harness routine: execute()}\\[-1pt]
\hspace{0.85em}\texttt{-> task\_completed()}\\[2pt]
\textit{Why it explains the link}\\[-1pt]
{\tiny$\bullet$}~\texttt{status=success permits}\\[-1pt]
\hspace{0.85em}\texttt{complete\_task()}};
\draw[arr] (steps.east) -- (cf.west);
\draw[dasharr] (impl.north) -- (cf.south);
\end{tikzpicture}
\end{figure}

\subsubsection{\textbf{Implementation Anchors: Localizing Runtime Evidence in Harness Artifacts}}
TraceSteps and their associated data-flow and control-flow links describe runtime events and cross-step relations, but they do not by themselves identify where the corresponding behavior is realized in the harness implementation.
HTIR therefore associates runtime evidence with \textit{implementation anchors}, which localize the editable harness artifacts that realize or govern the observed behavior.

The trace abstraction agent derives implementation anchors by jointly inspecting the trajectory and the harness artifacts.
It leverages available runtime signals, such as tool names, prompt template references, adapter invocations, validator events, and controller actions, to locate the corresponding implementation artifacts.
When a relation is realized across multiple components, the agent follows the previously constructed control-flow and data-flow links
to identify the harness artifacts that implements or constrains the relation.
Each anchor records three fields: an artifact reference, such as a repository path and line range, prompt or configuration identifier, tool-schema identifier, workflow-node identifier, or validator identifier; an anchor relation, such as emitting or executing a TraceStep, constructing a step input, or governing a control-flow transition; and supporting evidence for this alignment.

The example below grounds the S3–S4 data-flow link (shown earlier) in the prompt rendering artifact that exposes the S3 API documentation.

\begin{figure}[H]
\refstepcounter{figure}\label{fig:anchor_localization}
\centering
\scriptsize
\setlength{\fboxsep}{1pt}
\begin{tikzpicture}[
    box/.style={draw, rounded corners=1pt, align=left, inner sep=3pt, text width=0.98\columnwidth, font=\scriptsize},
    arr/.style={->, thick, purple!70}
]

\node[box] (anchor) at (0.50\columnwidth,0) {\textbf{Implementation anchor}\\[-1pt]
{\tiny$\bullet$}~\texttt{artifact reference:}\\[-1pt]
\hspace{0.85em}\texttt{prompts/react\_code\_agent/}\\[-1pt]
\hspace{0.85em}\texttt{instructions.txt}\\[-1pt]
\hspace{0.85em}\texttt{code/simplified/react\_code\_agent.py}\\[-1pt]
\hspace{0.85em}\texttt{SimplifiedReActCodeAgent.initialize()}\\[1pt]
{\tiny$\bullet$}~\texttt{anchor relation: constructs S4 input}\\[1pt]
\textit{Relevant prompt/template lines}\\[-1pt]
\hspace{0.85em}\texttt{instructions.txt:}\\[-1pt]
\hspace{1.70em}\texttt{``interact with app(s) ... using APIs''}\\[-1pt]
\hspace{1.70em}\texttt{show\_api\_doc(app\_name, api\_name)}\\[-1pt]
\hspace{1.70em}\texttt{``Each code execution will produce}\\[-1pt]
\hspace{1.70em}\texttt{an output ...''; Task: \{\{ instruction \}\}}\\[-1pt]
\hspace{0.85em}\texttt{initialize():}\\[-1pt]
\hspace{1.70em}\texttt{template.render(instruction, main\_user,}\\[-1pt]
\hspace{1.70em}\texttt{app\_descriptions) -> self.messages}\\[1pt]
{\tiny$\bullet$}~\texttt{supporting evidence: ...}};

\end{tikzpicture}
\end{figure}

\subsubsection{\textbf{Harness-layer Responsibility Mapping}}
After TraceSteps and their associated links are connected to harness artifacts through implementation anchors, HTIR maps the anchored evidence around each TraceStep to ETCLOVG responsibility layers.
This mapping is used to consolidate recurring flaw records, select scoped repair operators, and analyze which harness responsibilities are implicated by different benchmark failures.
It is trace-specific rather than a static file-level classification: the trace abstraction agent takes the TraceStep's information, associated links, and implementation anchors as input, and compares them with the responsibility criteria of the ETCLOVG layers in Table~\ref{tab:htir}.
A TraceStep can map to multiple layers when the anchored runtime evidence involves multiple harness responsibilities.

Using the diagnostic information above, we analyze the layer mapping for S6.
S6 invokes \texttt{complete\_task()} after the S5--S6 transition is enabled by \texttt{status=success}.
The supporting S3--S4 data-flow link provides contextual diagnostic evidence: the S3 API documentation specifies the required \texttt{user\_email} field, but the S4 payment request body omits it.
Together, these signals indicate that S6 finalizes the task despite an unresolved upstream request-construction error and the absence of expected artifact/state effects, leading to its mapping to the Lifecycle and Verification layers.
This discussion focuses on the layer mapping for S6; the complete flaw record for this case also involves S5 Observability evidence showing that API errors are swallowed and artifact/state effect evidence is not exposed to the harness.

\subsection{Harness Flaw Diagnosis}
\label{subsec:harness_flaw_diagnosis}

\subsubsection{\textbf{Failure Attribution}}

Given a failed execution trace, failure attribution identifies which TraceSteps are responsible for the failure, determines the root cause, and links it to specific harness layers.

The diagnosis agent proceeds with the task in four steps.
First, symptom localization identifies what failed by examining the external evaluation result together with the final TraceStep's diagnostic evidence, including the TraceStep details, associated links, implementation anchors, and mapped harness layers.
Second, evidence backtracking follows the data-flow and control-flow links from the final TraceStep to earlier TraceSteps and produces a ranked set of candidate responsible TraceSteps.
This stage is deliberately broad: a candidate step can be suspicious because it is plausibly connected to the observed failure through data propagation, control transition, artifact/state effect, or an anchored harness mechanism.
Third, candidate adjudication inspects each candidate's diagnostic evidence
to decide whether the step formed, propagated, failed to expose, failed to constrain, or failed to validate the information or state change relevant to the failure. It then selects the final responsible TraceSteps.
Fourth, layer assignment attributes the failure to the harness layers already mapped to the final responsible TraceSteps.

For each failed trace, {\tool} produces a structured \textit{diagnosis record} with three parts: (a) \textit{Attribution result} identifies the responsible TraceSteps and provides a concise root-cause explanation.
(b) \textit{Failure signal} records the observed failure symptom and related diagnostic evidence supporting this attribution.
(c) \textit{Harness diagnosis} includes one or more implicated ETCLOVG harness layers and a concise description of the harness flaw.

\subsubsection{\textbf{Harness Flaw Consolidation}}

Single failed executions can be accidental. {\tool} therefore consolidates diagnosis records across task executions before modifying the harness.
The diagnosis agent first indexes diagnosis records by their implicated harness layers, and treats records with overlapping layers as merge candidates.
It then merges candidates whose root cause explanations and supporting evidence indicate the same recurring harness flaw.

Each resulting \textit{flaw record} represents a recurring harness flaw pattern and is organized into two groups of fields:
(a)\textit{ Flaw summary} describes the recurring harness flaw. It includes a flaw identifier, one or more implicated harness layers,
and a common root cause summary.
(b)\textit{ Supporting diagnosis records} summarizes the empirical basis for the flaw. It includes the representative diagnosis records and the shared diagnostic basis used to merge them, such as common failure signals, responsible TraceSteps, and root cause rationales across the grouped diagnosis records.

\subsection{Scoped Repair}
\label{subsec:scoped_repair}
After flaw consolidation, {\tool} must decide what kind of harness change is appropriate for each recurring flaw. This stage is deliberately constrained because the harness artifacts implicated by failure attribution often belong to core runtime logic, such as orchestration and context construction. Allowing the repair agent to freely modify such artifacts, as in self-improvement coding agent systems~\cite{godelagent2024,sica2025,dgm2025,huxleygodel2026}, would create a high risk of unstable edits, architectural breakage, or broad regressions.

Therefore, {\tool} applies repair operators, derived from our motivational study and summarized in Table~\ref{tab:operators}, as scoped repair operations that specify what harness mechanisms may be changed for diagnosed flaws.
{\tool} first maps each flaw record to applicable repair operators, and instantiates the selected operators using the flaw record's supporting diagnosis records to derive flaw-specific repair specifications.

\subsubsection{\textbf{Mapping Flaw to Repair Operators}}

The input to this stage is a flaw record, and {\tool} maps it to applicable repair operators.
The repair agent first uses the implicated harness layers in the flaw record to retrieve candidate operator groups from Table~\ref{tab:operators}.
It then filters these candidates according to the flaw record's common root cause and diagnostic evidence.
From the selected operators, it designates a primary operator based on how directly its editable mechanism targets the implicated TraceStep and whether its expected effect directly satisfies the behavioral requirement implied by the root cause.
The remaining selected operators serve as auxiliary operators: they expose evidence, enforce constraints, or modify adjacent harness mechanisms needed for the primary repair to work.
The selected operators and their selection rationale are carried into the repair specification, allowing patch validation to check whether the generated diff follows the intended repair scope.

\subsubsection{\textbf{Patch Generation with Repair Specification}}

After repair operators have been selected for a flaw record, {\tool} instantiates them into a repair specification for candidate implementation. The purpose of the specification is to translate operator-level repair intent into a flaw-specific implementation contract, i.e., it binds the operator to the current flaw record and target harness. The resulting specification replaces abstract operator fields with flaw-specific and harness-specific constraints.
Each specification contains three groups of fields.

(a)\textit{ Target and scope} binds the specification to a concrete flaw record, its implicated harness layers, selected repair operators, and representative diagnosis records.

(b)\textit{ Edit constraints} bind the operators' editable resource template to concrete harness artifacts in the target implementation, while listing forbidden artifacts.

(c) \textit{Required behavior} translates the expected effect of the operators into concrete behavioral requirements.
It describes the harness behavior that must hold after applying the candidate change.

The example below sketches the repair specification instantiated from the consolidated flaw records for  task in \Cref{fig:htir_example}. 

\begin{figure}[H]
\centering
\scriptsize
\setlength{\fboxsep}{1pt}
\begin{tikzpicture}[
    rec/.style={draw, rounded corners=1pt, align=left, inner sep=3pt, text width=0.92\columnwidth, font=\scriptsize}
]
\node[rec] (r) at (0.48\columnwidth,0) {\textbf{Repair specification}\\[-1pt]
{\tiny$\bullet$}~\texttt{target and scope:}\\[-1pt]
\hspace{0.85em}{\tiny$\circ$}~\texttt{flaw record=completion accepted}\\[-1pt]
\hspace{1.70em}\texttt{without artifact/state effect}\\[-1pt]
\hspace{0.85em}{\tiny$\circ$}~\texttt{implicated layers=Lifecycle,}\\[-1pt]
\hspace{1.70em}\texttt{Verification, Observability}\\[-1pt]
\hspace{0.85em}{\tiny$\circ$}~\texttt{selected operators=verification-gated}\\[-1pt]
\hspace{1.70em}\texttt{finalization; effect-evidence}\\[-1pt]
\hspace{1.70em}\texttt{completion guarding;}\\[-1pt]
\hspace{1.70em}\texttt{API-error and state-delta logging}\\[-1pt]
\hspace{0.85em}{\tiny$\circ$}~\texttt{representative diagnoses=...}\\[2pt]
{\tiny$\bullet$}~\texttt{edit constraints:}\\[-1pt]
\hspace{0.85em}{\tiny$\circ$}~\texttt{allowed artifacts=appworld\_agent/core.py;}\\[-1pt]
\hspace{1.70em}\texttt{execution-result instrumentation}\\[-1pt]
\hspace{0.85em}{\tiny$\circ$}~\texttt{forbidden artifacts=task data/oracles,}\\[-1pt]
\hspace{1.70em}\texttt{validation/test labels, public APIs}\\[2pt]
{\tiny$\bullet$}~\texttt{required behavior:}\\[-1pt]
\hspace{0.85em}{\tiny$\circ$}~\texttt{expose API-error evidence}\\[-1pt]
\hspace{0.85em}{\tiny$\circ$}~\texttt{require artifact/state effect}\\[-1pt]
\hspace{1.70em}\texttt{evidence before complete\_task()}\\[-1pt]
\hspace{1.70em}\texttt{e.g., PaymentRequest records}};
\end{tikzpicture}
\end{figure}

\subsection{Patch Validation and Harness Memory}
\label{subsec:patch_validation_harness_memory}

With the candidate patch generated under the repair specification, {\tool} enters a validation stage that determines whether the patch should be accepted into the harness.

The validation agent first performs pre-validation checks on the candidate diff (between the generated patch and the original implementation) to determine whether it conforms to the repair specification and satisfies basic implementation correctness, such as syntax and static analysis checks.

It then performs validation set assessment by evaluating the candidate patch on a held-out validation set, checking whether it mitigates the target flaw while avoiding unacceptable regressions on tasks previously solved by the original harness.
A patch is accepted only if it achieves the target improvement and stays within the regression limit.

{\tool} records the outcome as harness repair memory.
Each memory record contains the flaw record, implicated harness layers, repair specification, diff summary, results from pre-validation check and validation set assessment, the validation tasks that regressed, and a natural language description of the task conditions under which the repair applies or does not apply.
Accepted records provide implementation examples for similar future flaw records. Rejected records explain whether the cause is pre‑validation failure, insufficient target improvement, or excessive regression, and help prevent future iterations from proposing the same ineffective change.

\section{Experimental Design}
\label{sec_Experiment_Design}

\subsection{Research Questions}

\textbf{RQ1: Effectiveness and Efficiency.} Does {\tool} achieve better task performance and token efficiency?

\textbf{RQ2: Failure diagnosis.} Does {\tool} produce effective failure diagnoses?

\textbf{RQ3: Ablation study.} Do {\tool}'s design choices all contribute to performance gains?

\textbf{RQ4: Cross-model transfer.} Do harness repairs derived from one task model transfer to other task models?

\subsection{Agents and Benchmarks}
We evaluate {\tool} on four representative benchmarks for agent harnesses, covering open-ended research QA, repository-level software repair, stateful application automation, and terminal-based command-line workflows.
For each benchmark, we select one initial harness $H_0$ as the starting point for {\tool} and the self-evolution baselines, and randomly partition the sampled tasks into training, validation, and held-out test sets.

For GAIA~\cite{gaia2023}, we use the open-deep-research harness~\cite{opendeepresearch2025} as $H_0$ and sample 150 of the 166 annotated developer questions with public ground truth, split into 60/30/60 tasks.
For SWE-Bench Verified~\cite{swebenchverified2024}, we use mini-swe-agent~\cite{minisweagent2024} as $H_0$ and sample 250 of the 500 instances, split into 100/50/100 tasks.
For AppWorld~\cite{appworld2024}, we use the official simplified ReAct code agent as $H_0$ and sample 225 of the 750 tasks, split into 90/45/90 tasks.
For Terminal-Bench 2.0 Verified~\cite{terminalbench2026}, we use the Harbor Terminus-2 terminal-agent harness~\cite{harborterminus2026} as $H_0$ and sample 85 of the 89 tasks, split into 34/17/34 tasks.

Following the standard benchmark practice, we measure performance by task completion rate (short for \texttt{TCR}).

\subsection{Baselines}
RQ1 compares {\tool} with two categories of baselines: human-designed harness baselines and self-evolution and repair baselines.

\textbf{Human-designed harness baselines.}
For each benchmark, we compare against representative human-designed harnesses: DeepResearchAgent~\cite{deepresearchagent2025} and MiroFlow~\cite{miroflow2025} for GAIA; OpenHands~\cite{openhands2025} and Trae-Agent~\cite{traeagent2025} for SWE-Bench Verified; FullCodeRefl, IPFunCall, and CUGA~\cite{cuga2025} for AppWorld; and OpenCode~\cite{opencode2024} and OpenHands~\cite{openhands2025} for Terminal-Bench 2.0 Verified.

\textbf{Self-evolution and repair baselines.}
We also compare with four representative approaches that adapt agent behavior from observed executions: \textbf{1) GEPA~\cite{agrawal2026gepa}}, a reflective prompt-evolution method with Pareto-frontier selection; \textbf{2) SCOPE~\cite{scope2025}}, a memory-guided prompt-evolution method; \textbf{3) ReCreate~\cite{hao2026recreate}}, which converts trajectories into reusable task-domain scaffolds and workflow guidance; and \textbf{4) Meta-Harness~\cite{metaharness2026}}, which optimizes harness code from prior harness versions, task scores, and execution traces.
For these baselines, we also record token consumption (in millions), short for \texttt{Tokens (M)}.

\subsection{Experimental Setup}

We use GPT-5 mini as the default model for running the experimental agents, our {\tool} and baselines.
We also experiment with four other LLMs for comparison and evaluate the cross-model transfer performance (in RQ4). Detailed model configurations are on our website.
Task performance values in Tables~\ref{tab:end_to_end}, \ref{tab:gates}, \ref{tab:repair_source_sensitivity}, and \ref{tab:model_transfer} are arithmetic means over three independent runs.

To answer RQ2, we construct a human-annotated diagnosis set from the harness failures on the test set, and this information is not used for harness repair or validation decisions.
For each benchmark, three annotators label 20 failed test trajectories with the responsible failure step, root cause, implementation anchor, implicated harness layer, and selected repair operators.
These labels serve as the gold standard for evaluation, and we report diagnostic \texttt{accuracy} across the annotated dimensions.
The full annotation and metric details are provided on our website.

To answer RQ3, we evaluate four ablations that remove or weaken key design choices in {\tool}: \textbf{1) Prompt-only repair}, which applies only prompt updates and disables runtime harness changes; \textbf{2) w/o trace-grounded diagnosis}, which removes Harness Flaw Diagnosis (\Cref{subsec:harness_flaw_diagnosis}) and derives repair context from raw trajectory summaries; \textbf{3) w/o scoped repair operators}, which replaces Scoped Repair (\Cref{subsec:scoped_repair}) with free-form harness editing; and \textbf{4) w/o regression-aware acceptance}, which removes validation-set acceptance checks (\Cref{subsec:patch_validation_harness_memory}) for target-flaw reduction and newly introduced regressions.

To answer RQ4, we use the repaired GAIA harness selected with GPT-5 mini and directly reuse it with four additional task models on the same held-out GAIA test split: Qwen3.5 Plus, DeepSeek V3.2, Gemini 3 Pro, and Claude Sonnet 4.5.
Apart from minimal model-provider compatibility handling required to execute the target APIs, no target-model-specific harness repair is performed for these transfer runs.

\section{Results and Analysis}
\label{sec_results}

\subsection{Effectiveness and Efficiency (RQ1)}

\begin{table}[!t]
\centering
\vspace{0.05in}
\caption{Performance over $H_0$ (RQ1).
}
\label{tab:repair_source_sensitivity}
\scriptsize
\setlength{\tabcolsep}{3pt}
\renewcommand{\arraystretch}{1.02}
\begin{tabular}{@{}llrrr@{}}
\toprule
\textbf{Bench.} & \textbf{LLM used} & \textbf{$H_0$} & \textbf{$H_0$+\tool{}} & \textbf{$\Delta$} \\
\midrule
\multirow[c]{5}{*}{GAIA}
 & GPT-5 mini & 43.3 & 61.7 & +18.4 \\
 & Claude Sonnet 4.5 & 66.7 & 84.4 & +17.7 \\
 & DeepSeek V3.2 & 58.9 & 75.6 & +16.7 \\
 & Qwen3.5 Plus & 63.3 & 78.9 & +15.6 \\
 & Gemini 3 Pro & 69.4 & 83.9 & +14.5 \\
\midrule
\multirow[c]{5}{*}{SWE}
 & GPT-5 mini & 45.3 & 57.3 & +12.0 \\
 & Claude Sonnet 4.5 & 61.3 & 71.7 & +10.4 \\
 & DeepSeek V3.2 & 58.7 & 69.3 & +10.6 \\
 & Qwen3.5 Plus & 62.7 & 72.3 & +9.6 \\
 & Gemini 3 Pro & 60.3 & 68.7 & +8.4 \\
\midrule
\multirow[c]{5}{*}{App.}
 & GPT-5 mini & 36.7 & 43.0 & +6.3 \\
 & Claude Sonnet 4.5 & 61.1 & 70.4 & +9.3 \\
 & DeepSeek V3.2 & 56.3 & 62.2 & +5.9 \\
 & Qwen3.5 Plus & 53.3 & 61.9 & +8.6 \\
 & Gemini 3 Pro & 50.7 & 59.6 & +8.9 \\
\midrule
\multirow[c]{5}{*}{TB2}
 & GPT-5 mini & 17.6 & 26.5 & +8.9 \\
 & Claude Sonnet 4.5 & 34.3 & 47.1 & +12.7 \\
 & DeepSeek V3.2 & 31.4 & 39.2 & +7.9 \\
 & Qwen3.5 Plus & 42.2 & 52.9 & +10.7 \\
 & Gemini 3 Pro & 46.1 & 55.9 & +9.8 \\
\bottomrule
\end{tabular}
\vspace{-0.1in}
\end{table}

\begin{table}[!t]
\centering
\vspace{-0.1in}
\caption{Performance comparison (RQ1). 
}
\label{tab:end_to_end}
\scriptsize
\setlength{\tabcolsep}{1.4pt}
\renewcommand{\arraystretch}{1.03}
\begin{tabular}{p{0.6cm}<{\centering}  p{1.5cm}<{\centering}  p{2.6cm}<{\centering}  p{1.6cm}<{\centering}  p{1.8cm}<{\centering} }
\toprule
\textbf{Bench.} & \textbf{Role} & \textbf{Method / harness} & \textbf{TCR} & \textbf{Tokens (M)} \\
\midrule

\multirow[c]{8}{0.09\columnwidth}{GAIA}
 & Selected $H_0$ & open-deep-research & 43.3 {\scriptsize(+18.4)} & / \\
\cmidrule(l{0.45em}r{-0.35em}){2-5}
 & \multirow[c]{2}{0.14\columnwidth}{Human-designed} & DeepResearchAgent & 52.8 {\scriptsize(+8.9)} & / \\
 & & MiroFlow & \uline{58.3} {\scriptsize(+3.4)} & / \\
\cmidrule(l{0.45em}r{-0.35em}){2-5}
 & \multirow[c]{5}{0.14\columnwidth}{Evolve/repair from $H_0$} & $H_0$ + GEPA & 46.7 {\scriptsize(+15.0)} & 32.6 {\scriptsize(-43.4)} \\
 & & $H_0$ + SCOPE & 45.0 {\scriptsize(+16.7)} & 26.9 {\scriptsize(-53.3)} \\
 & & $H_0$ + ReCreate & 52.8 {\scriptsize(+8.9)} & 67.9 {\scriptsize(+17.9)} \\
 & & $H_0$ + Meta-Harness & 56.7 {\scriptsize(+5.0)} & 94.2 {\scriptsize(+63.5)} \\
 & & \textbf{$H_0$ + \tool{}} & \textbf{61.7} & 57.6 \\
\midrule
\multirow[c]{8}{0.09\columnwidth}{SWE}
 & Selected $H_0$ & mini-swe-agent & 45.3 {\scriptsize(+12.0)} & / \\
\cmidrule(l{0.45em}r{-0.35em}){2-5}
 & \multirow[c]{2}{0.14\columnwidth}{Human-designed} & OpenHands & 47.3 {\scriptsize(+10.0)} & / \\
 & & Trae-Agent & 48.7 {\scriptsize(+8.6)} & / \\
\cmidrule(l{0.45em}r{-0.35em}){2-5}
 & \multirow[c]{5}{0.14\columnwidth}{Evolve/repair from $H_0$} & $H_0$ + GEPA & 46.7 {\scriptsize(+10.6)} & 28.1 {\scriptsize(-42.5)} \\
 & & $H_0$ + SCOPE & 48.3 {\scriptsize(+9.0)} & 25.6 {\scriptsize(-47.6)} \\
 & & $H_0$ + ReCreate & 51.7 {\scriptsize(+5.6)} & 57.2 {\scriptsize(+17.0)} \\
 & & $H_0$ + Meta-Harness & \uline{54.7} {\scriptsize(+2.6)} & 82.9 {\scriptsize(+69.5)} \\
 & & \textbf{$H_0$ + \tool{}} & \textbf{57.3} & 48.9 \\
\midrule
\multirow[c]{9}{0.09\columnwidth}{App.}
 & Selected $H_0$ & ReAct & 36.7 {\scriptsize(+6.3)} & / \\
\cmidrule{2-5}
 & \multirow[c]{3}{0.14\columnwidth}{Human-designed} & FullCodeRefl & 35.6 {\scriptsize(+7.4)} & / \\
 & & IPFunCall & 38.1 {\scriptsize(+4.9)} & / \\
 & & CUGA & \uline{41.1} {\scriptsize(+1.9)} & / \\
\cmidrule(l{0.45em}r{-0.35em}){2-5}
 & \multirow[c]{5}{0.14\columnwidth}{Evolve/repair from $H_0$} & $H_0$ + GEPA & 37.4 {\scriptsize(+5.6)} & 22.7 {\scriptsize(-39.0)} \\
 & & $H_0$ + SCOPE & 38.9 {\scriptsize(+4.1)} & 18.8 {\scriptsize(-49.5)} \\
 & & $H_0$ + ReCreate & 39.3 {\scriptsize(+3.7)} & 44.1 {\scriptsize(+18.5)} \\
 & & $H_0$ + Meta-Harness & \uline{40.4} {\scriptsize(+2.6)} & 74.6 {\scriptsize(+100.5)} \\
 & & \textbf{$H_0$ + \tool{}} & \textbf{43.0} & 37.2 \\
\midrule
\multirow[c]{8}{0.09\columnwidth}{TB2}
 & Selected $H_0$ & Harbor Terminus-2 & 17.6 {\scriptsize(+8.9)} & / \\
\cmidrule(l{0.45em}r{-0.35em}){2-5}
 & \multirow[c]{2}{0.14\columnwidth}{Human-designed} & OpenCode & 22.5 {\scriptsize(+4.0)} & / \\
 & & OpenHands & 18.6 {\scriptsize(+7.9)} & / \\
\cmidrule(l{0.45em}r{-0.35em}){2-5}
 & \multirow[c]{5}{0.14\columnwidth}{Evolve/repair from $H_0$} & $H_0$ + GEPA & 19.6 {\scriptsize(+6.9)} & 18.6 {\scriptsize(-44.5)} \\
 & & $H_0$ + SCOPE & 20.6 {\scriptsize(+5.9)} & 18.1 {\scriptsize(-46.0)} \\
 & & $H_0$ + ReCreate & 21.6 {\scriptsize(+4.9)} & 38.8 {\scriptsize(+15.8)} \\
 & & $H_0$ + Meta-Harness & \uline{23.5} {\scriptsize(+3.0)} & 66.8 {\scriptsize(+99.4)} \\
 & & \textbf{$H_0$ + \tool{}} & \textbf{26.5} & 33.5 \\
\bottomrule
\end{tabular}
\par\smallskip
\noindent\parbox{0.96\columnwidth}{\scriptsize \textit{Note.} In TCR, parentheses show \tool{}'s percentage-point gain over the row. In Tokens, parentheses show relative token change vs. \tool{}; / denotes not applicable.}
\vspace{-0.15in}
\end{table}

\Cref{tab:repair_source_sensitivity} compares
{\tool} with the initial harness $H_0$ across four benchmarks and five LLMs.
{\tool} consistently improves over $H_0$ in all settings, with an average gain of 11.1\%.
The largest improvements appear on GAIA, ranging from 14.5\% to 18.4\%, while even the smallest gains on AppWorld remain between 5.9\% and 9.3\%.
The absolute scores differ across LLMs, but the improvement introduced is relatively model consistent: the average gain for each LLM ranges from 10.3\% to 12.5\%.

In \Cref{tab:end_to_end}, we fix the backbone model to GPT-5 mini, one of the five models evaluated above, and compare {\tool} with all baselines end to end.
{\tool} outperforms the human-designed harness by 6.3\% on average, with improvements between 1.9\% and 10.0\%, while requiring no manual harness design effort.
Among automated self-evolution and repair baselines initialized from the same $H_0$, {\tool} achieves the best performance, improving over these baselines by 6.9\% on average, with gains ranging from 2.6\% to 16.7\%.
Even against Meta-Harness, the strongest automated baseline, {\tool} remains 2.6\% to 5.0\% higher while using substantially fewer tokens.
Specifically, Meta-Harness consumes 63.5\% to 100.5\% more offline evolving/repair tokens than {\tool}, indicating that our approach improves both effectiveness and token efficiency.

To evaluate the statistical significance of these improvements, we conduct one-sided paired sign tests using the repeated-run outcomes across all benchmarks.
For the baseline comparisons in Table~\ref{tab:end_to_end}, the $p$-values range from $2.4\times 10^{-4}$ to $4.9\times 10^{-4}$, all below 0.001, showing that {\tool}'s improvements are statistically significant.

\subsection{Failure Diagnosis (RQ2)}

\begin{table}[!t]
\vspace{-0.1in}
\centering
\caption{Failure diagnosis evaluation (RQ2). 
}
\label{tab:diagnosis_quality}
\footnotesize
\setlength{\tabcolsep}{2.2pt}
\renewcommand{\arraystretch}{1.05}
\begin{tabular}{lccccc}
\toprule
\textbf{Input representation} & \textbf{Step} & \textbf{Cause} & \textbf{Anchor} & \textbf{Layer} & \textbf{Operator} \\
\midrule
Raw trace & 55.0 & 53.8 & 50.0 & 58.4 & 51.3 \\
Raw + data-flow & 70.0 & 68.8 & 65.0 & 73.2 & 66.3 \\
Raw + data/control & 77.5 & 75.0 & 72.5 & 79.4 & 73.8 \\
\textbf{Full HTIR} & \textbf{85.0} & \textbf{83.8} & \textbf{81.3} & \textbf{86.2} & \textbf{82.5} \\
\bottomrule
\end{tabular}
\end{table}

\Cref{tab:diagnosis_quality} shows that Full HTIR aligns well with human-annotated gold labels, achieving 85.0\% step accuracy, 81.3\% implementation anchor accuracy, 86.2\% harness-layer macro-F1, and 82.5\% repair-operator accuracy.
For fine-grained step-level localization, this is substantially above existing techniques that typically achieve around 45--52\% step accuracy \cite{2026fromflatlogsCHIEF,ma2025doverinterventiondrivenautodebugging}, raising the diagnosis to a level sufficient for targeted repair rather than blind optimization.
Compared with raw traces, adding data-flow and control-flow links consistently improves the diagnosis metrics.
These results indicate that {\tool}'s gains are driven by structured, trace-grounded diagnosis.

\begin{table}[!t]
\centering
\caption{Ablation study (RQ3).}
\label{tab:gates}
\small
\resizebox{\columnwidth}{!}{%
\begin{tabular}{lcccc}
\toprule
\textbf{Variant} & \textbf{GAIA} & \textbf{SWE} & \textbf{AppWorld} & \textbf{TB2} \\
\midrule
$H_0$ & 43.3 & 45.3 & 36.7 & 17.6 \\
Prompt-only repair & 50.6 & 48.3 & 37.4 & 18.6 \\
w/o trace-grounded diagnosis & 51.1 & 50.7 & 38.1 & 21.6 \\
w/o scoped repair operators & 50.6 & 49.3 & 37.4 & 18.6 \\
w/o regression-aware acceptance & 55.6 & 53.3 & 39.3 & 24.5 \\
\textbf{Full \tool{}} & \textbf{61.7} & \textbf{57.3} & \textbf{43.0} & \textbf{26.5} \\
\bottomrule
\end{tabular}}
\end{table}

\subsection{Ablation Study (RQ3)}

As shown in \Cref{tab:gates}, all ablations reduce performance relative to full {\tool}, showing that each design choice contributes to the final result.
The largest degradation is observed when repair is limited to prompt updates or when scoped repair operators are removed, indicating that effective harness repair depends on targeted changes to runtime mechanisms.

\subsection{Cross-model Transfer (RQ4)}

\begin{table}[!t]
\centering
\caption{Cross-model transfer on GAIA (RQ4)
}
\label{tab:model_transfer}
\scriptsize
\setlength{\tabcolsep}{2.0pt}
\renewcommand{\arraystretch}{1.05}
\begin{threeparttable}
\begin{tabularx}{\columnwidth}{>{\raggedright\arraybackslash}X>{\raggedleft\arraybackslash}m{0.17\columnwidth}>{\raggedleft\arraybackslash}m{0.29\columnwidth}>{\raggedleft\arraybackslash}m{0.12\columnwidth}}
\toprule
\textbf{Model} & \textbf{$H_0$} & \textbf{$H_0$ + \tool{}} & \textbf{$\Delta$} \\
\midrule
GPT-5 mini (repair source) & 43.3 & 61.7 & +18.4 \\
\midrule
Claude Sonnet 4.5 & 66.7 & 72.2 & +5.5 \\
DeepSeek V3.2 & 58.9 & 66.7 & +7.8 \\
Qwen3.5 Plus & 63.3 & 72.8 & +9.5 \\
Gemini 3 Pro & 69.4 & 78.3 & +8.9 \\
\bottomrule
\end{tabularx}
\end{threeparttable}
\vspace{-0.1in}
\end{table}

\Cref{tab:model_transfer} reports whether the GAIA harness repaired with GPT-5 mini transfers to other models without target-specific trajectory analysis.
The repaired harness improves the repair-source model by 18.4\%, and also improves all four target models, yielding consistent transfer gains of 5.5\% to 9.5\%.
These gains are smaller than the source-model gain but consistently positive, suggesting that the repairs address harness-level failures shared across models.

\section{Discussion}
\label{sec_discussion}

\begin{table}[!t]
\vspace{0.05in}
\centering
\caption{Harness layers repaired by {\tool} and baselines. \CIRCLE\ = primary edit target, \LEFTcircle\ = partial or indirect edit target, \Circle\ = not edited.
}
\label{tab:harnessfix_edits}
\scriptsize
\setlength{\tabcolsep}{2.1pt}
\renewcommand{\arraystretch}{1.05}
\begin{tabular}{@{}lcccc||cccc@{}}
\toprule
\multirow{2}{*}{\textbf{Layer}} & \multicolumn{4}{c||}{\textbf{\tool{} repairs}} & \multicolumn{4}{c}{\textbf{Baselines}} \\
\cmidrule(lr){2-5}\cmidrule(l){6-9}
 & \textbf{GAIA} & \textbf{SWE} & \textbf{App.} & \textbf{TB2} & \textbf{GEPA} & \textbf{SCOPE} & \textbf{ReC.} & \textbf{Meta-H.} \\
\midrule
Execution & \LEFTcircle & \CIRCLE & \Circle & \Circle & \Circle & \Circle & \Circle & \CIRCLE \\
Tool Interface & \CIRCLE & \LEFTcircle & \LEFTcircle & \Circle & \Circle & \Circle & \LEFTcircle & \CIRCLE \\
Context/Memory & \LEFTcircle & \CIRCLE & \CIRCLE & \CIRCLE & \CIRCLE & \CIRCLE & \CIRCLE & \CIRCLE \\
Lifecycle & \CIRCLE & \CIRCLE & \LEFTcircle & \CIRCLE & \Circle & \Circle & \Circle & \CIRCLE \\
Observability & \CIRCLE & \LEFTcircle & \Circle & \Circle & \Circle & \Circle & \Circle & \LEFTcircle \\
Verification & \Circle & \CIRCLE & \CIRCLE & \CIRCLE & \Circle & \Circle & \Circle & \CIRCLE \\
Governance & \Circle & \CIRCLE & \LEFTcircle & \CIRCLE & \Circle & \Circle & \Circle & \LEFTcircle \\
\midrule
\# layers touched & 5 & 7 & 5 & 4 & 1 & 1 & 2 & 7 \\
\bottomrule
\end{tabular}
\par\vspace{1pt}
\noindent\parbox{\columnwidth}{\scriptsize Baseline columns summarize method-level edit scope. ReC. = ReCreate; Meta-H. = Meta-Harness.}
\vspace{-0.1in}

\end{table}

Table \ref{tab:harnessfix_edits} shows that most baselines modify only a narrow slice of the harness: GEPA and SCOPE touch one layer, and ReCreate touches two, making them less suitable for failures spanning multiple harness responsibilities.
Meta-Harness can touch all seven layers, but its optimization-based search is untargeted and therefore incurs much higher token costs as shown in \Cref{tab:end_to_end}.
In contrast, {\tool} uses trace-grounded and diagnosis-driven framework to perform scoped edit, covering diverse harness flaws without exhaustive search.

Table \ref{tab:harnessfix_edits} also demonstrates the concrete harness repair conducted by {\tool}.
In open-ended research QA (GAIA), flaws often concentrate in Tool Interface and Observability, such as missing API configuration, unsupported document formats, fragile media conversion, and traces that do not preserve enough evidence for later answer synthesis. In repository-level software repair (SWE-Bench Verified), many failures involve the interface between tool execution, context construction, and verification evidence: the agent must issue shell commands and file edits, observe the right test output, and submit a patch under benchmark-specific constraints.

Stateful application automation (AppWorld), where tasks require API calls or generated code to produce persistent application-state effects, exposes a different pattern. 
Its failures often combine Tool Interface, Lifecycle, Verification and  Governance concerns because the harness must validate API arguments, expose API errors, track artifact/state effects, and avoid accepting completion before the required side effect occurs. 
Terminal-style command-line tasks (Terminal-Bench 2.0 Verified) place more pressure on Context and Memory, Lifecycle, Verification, and Governance concerns: the harness must preserve terminal context, manage command-line workflow state, prefer benchmark-provided tests over ad hoc checks, and avoid invalid terminal actions.

These patterns suggest that harness repair is not only a prompt optimization problem. Different benchmarks require different runtime mechanisms, and a repair often has cross-layer implications. For example, a lifecycle repair that preserves partial work can also change the evidence available to verification and the conditions under which finalization should be accepted.
Because {\tool}'s diagnosis aggregates the implicated layers around each recurring flaw, its repair specification and validation checks can account for these coupled effects before accepting a candidate harness change.

\section{Related Work}
\label{sec_related}

\textbf{Agent Harnesses.}
Recent work treats the agent harness as a first-class engineering object rather than incidental glue code. 
Survey and practice-oriented work characterizes harness responsibilities and runtime surfaces~\cite{li2026agentharness,openharnessengineering2026,anthropicharnesses2025}.
A second line studies harness specification and composition: such as natural-language agent harnesses~\cite{nlah2026} and AgentFlow~\cite{agentflow2026}.
A third line automates harness optimization or evolution, 
including AutoHarness~\cite{autoharness2026}, Meta-Harness~\cite{metaharness2026}, and Agentic Harness Engineering~\cite{ahe2026}.
{\tool} builds on this view, but uses failed trajectories for diagnostic repair.

\textbf{Self-improving Agents.}
Self-improving agent work spans multiple repair surfaces.
Agent-design and workflow-search methods optimize agent structures, workflows, or multi-agent architectures~\cite{adas2025,aflow2024,autoflow2024,agentsquare2024,maas2025}.
Tool- and experience-oriented methods distill prior interactions into reusable tools, skills, workflows, memories, prompts, or domain agents~\cite{latm2023,creator2023,toolgen2025,trace2skill2026,skillforge2026,hao2026recreate,agentworkflowmemory2024,agentkb2025,amem2025,meminsight2025,ace2026,agrawal2026gepa}.
Self-modification and evolutionary methods recursively modify agent code, decision modules, collaboration networks, or development protocols~\cite{godelagent2024,sica2025,dgm2025,huxleygodel2026,groupagents2026,evomac2024,autogenesis2026,evotest2025}.
Verification-driven methods use rubric-guided checks or test-time learning to improve inference-time behavior~\cite{deepverifier2026,evotest2025}.
In contrast, {\tool} repairs the harness from failed trajectories: it attributes failures to responsible TraceSteps and harness layers, then applies scoped harness changes, rather than evolving the whole agent, merely reusing experience, or only strengthening inference-time verification.

\textbf{Trace Analysis, Failure Attribution and Agent Operations.}  
Wang et al. \cite{wang2026surveytrajectory} conducted the first survey about LLM agent trajectory analysis.
Some methods focus on representing or restructuring trajectories, such as SWE-TRACE ~\cite{swetrace2026} and VCC \cite{vcc2026}.
Others analyze how failures emerge over time ~\cite{beyondresolution2026,horizon2026}.
FAMAS \cite{ge2025introducingfailureautomaticallyattributing}, AgenTracer \cite{agentracer2025}, AgentFixer \cite{agentfixer2026} and DoVer \cite{ma2025doverinterventiondrivenautodebugging} further study failure attribution and root-cause diagnosis.
These works show the value of trajectory-level diagnosis; {\tool} further closes the loop by using diagnosis to implement harness changes.

\section{Conclusion}
\label{sec_conclusion}

As LLM-based agents are applied to increasingly complex tasks, the harness around the base model becomes an important part of system reliability. 
We propose {\tool}, which builds HTIR to organize heterogeneous trajectories and harness artifacts into step-level evidence, attributes failures to responsible steps and harness layers, and generates harness patches guided by flaw-specific repair specifications. 
As an early effort to study agent harness reliability, this work highlights the value of aligning runtime trajectories with harness implementations for diagnosis-aware repair, offering a new perspective on improving the reliability of LLM agents.

\bibliographystyle{IEEEtran}
\bibliography{reference,reference_from_survey}

\end{document}